\documentclass[twocolumn,showpacs,preprintnumbers,amsmath,amssymb]{revtex4}

\usepackage{graphicx}
\usepackage{dcolumn}
\usepackage{bm}

\begin{document}
\title{Spin blockade in a charge-switchable molecular magnet}

\author{C. Romeike}
 \email{romeike@physik.rwth-aachen.de}
\author{M. R. Wegewijs}
\author{H. Schoeller}

\affiliation{
Institut f\"ur Theoretische Physik A, RWTH Aachen,
 \\ 52056 Aachen,  Germany }

\date{\today}
\begin{abstract}
We consider the effect of adding electrons to a single molecule on its
magnetic properties
and the resulting transport fingerprints.
We analyze a generic model for a metal-organic complex consisting of
orbitals with different Coulomb repulsions.
We find that by modulating the charge of the molecule by a single
electron the total
spin can be switched from zero to the maximal value supported by the added
electrons, $S=3/2$. The Nagaoka mechanism is responsible for this charge-sensitivity of the
molecular spin. It is shown that fingerprints of these maximal spin states, either as
groundstates or low-lying excitations, can be experimentally observed
in current-spectroscopy.
as either spin blockade at low bias voltage
or negative differential conductance and complete current
suppression at finite bias.
\end{abstract}

\maketitle

\begin{figure}
\includegraphics[scale =0.4,angle=-90]{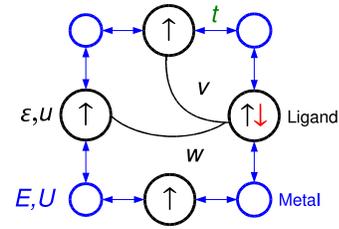}
\caption{\label{fig:geometry} Grid molecule: small/large circles
  represent metal ion/ligand orbitals.}
\end{figure}

\textbf{Introduction ---} Recent experiments on metal-organic \textit{grid complexes},
consisting of rationally designed ligands and metal ions as building
units have exhibited interesting electrochemical
~\cite{ruben04,zhao00} and magnetic ~\cite{waldmann02,waldmann04,guidi04} properties.
By self-assembly the  metal ions and ligands 
arrange in a rigid, highly symmetric grid.
 Due to their different nature, electron orbitals can often be
 roughly attributed either to the metal-ions or the ligands.
 Typically the orbitals on the (organic) ligands have  $\pi$ symmetry whereas the $d$ metal
orbitals split in an e.g. octahedral ligand field into subshells with local $\pi$ and
$\sigma$ symmetry. In the case of a fully occupied $d_{\pi}$ subshell 
tunneling between ligand $\pi$ orbitals and metal $d_{\sigma}$ orbitals
is weak due to their different symmetry.
 Such a separation into  metal ion and  ligand units
 has been used successfully to describe the
low-temperature intramolecular spin coupling of Co-[$2 \times 2$] grids
~\cite{waldmann97,waldmann_pre} and Mn-[$3 \times 3$] grids
~\cite{waldmann04,guidi04} for a fixed charge state as well as the
electrochemical properties of (Mn, Fe,
 Co, Zn)-[$2\times 2$]~\cite{ruben03a} and Mn-[$3 \times 3$] grids
~\cite{zhao00}. For poly-pyridine complexes it is well-known~\cite{vlcek82} that
ligands as  well as metal ions can be  reduced.
Which type is preferred depends on chemical details which can be
controlled, mainly by substitution  of metal ions and
 changing the ligand.
Here we analyze a phenomenological low temperature model for a [$2
\times 2$] grid molecule consisting of
 four ions and four ligands, Fig.~\ref{fig:geometry}. For this
 particular structure we show that (i)
 the molecular spin can be switched by the charge
and 
(ii) the spin-splitting appears in tunneling spectroscopy. 
 The well known Nagaoka
mechanism~\cite{nagaoka66} becomes effective for certain numbers of
added electrons. For strong onsite interaction the 
delocalization of an extra hole/electron relative to half-filling
(favoring a fully
polarized background of all other electrons) dominates over an antiferromagnetic
superexchange. 
In the context of band-magnetism the relevance of this mechanism is
limited due to its lattice type dependence~\cite{tasaki98} and its  strong charge sensitivity. Only for a single  additional 
electron or hole relative to  a half-filled band, the
spin-polarization effect can be guaranteed.
 In small single-molecule devices these obstacles can be overcome.
Firstly, the advanced rational design of supramolecular structures allows 
complex ``lattice'' types to be realized~\cite{lehn95, ruben04}. Secondly,
due to the Coulomb blockade effect one can modulate the total charge of a
molecule by a \textit{single} electron~\cite{park00, park02,liang02}.
In the case of sufficiently strong short-range interaction on the ligands (relative to
the ligand-ion tunneling)  the ground state spin is maximally increased 
 from $S=0$ to the maximal value supported by 3 (or 5) extra
electrons on the molecule, $S=3/2$.  
The magnetic properties of the molecule can thus be controlled
electrically~\cite{arita02,huai03}. 
In single electron tunneling transport this leads to spin blockade at low bias
voltage. Even for a low-lying maximal spin excitation
 negative differential conductance (NDC) effects and  complete
current suppression at finite bias voltage occur. 
Similar models with two types of electron orbitals have been studied for the description of the
neutral-ionic transition~\cite{torrence81} in organic crystals,  in
the context of ferroelectrics and superconductivity in
transition-metal oxides~\cite{ishihara94} and recently in the context of
exotic Kondo effects due to dynamical symmetries in multi quantum-dot systems~\cite{kuzmenko04,kikoin02}.
\\

\textbf{Model---}
We consider a model of a grid-complex with four metal  and four ligand
sites and one orbital per site (Fig.~\ref{fig:geometry}). For simplicity, we assume that the
metal ion can only be occupied virtually. The strong ligand
field separates electron accepting $d$ orbitals energetically from
ligand orbitals. Additionally, Coulomb repulsion on the ions is
typically much stronger than on the ligands. The following Hamiltonian  captures the features of the electronic
degrees  of freedom:
\begin{eqnarray}
  \label{eq:ham}
  H_{\mathsf{mol}} &=&  H_{\mathsf{T}} + H_{ \mathsf{L}} + H_{ \mathsf{M}},\\
  \label{eq:ham_tun}
  H_{\mathsf{T}} &=&  \sum_{\langle i,j\rangle}\sum_{\sigma} t\, A^{\dag}_{i \sigma} a_{j\sigma} + h.c.\\
  \label{eq:ham_lig}
  H_{ \mathsf{L}} &=& \sum_{j=1}^{4}( \epsilon n_{j} + u n_{j
    \uparrow} n_{j \downarrow} + v n_{j} n_{j+1} )  \nonumber \\
  &+& w \sum_{j=1}^{2}  n_{j} n_{j+2}  \\
  \label{eq:ham_met}
  H_{\mathsf{M}} &=& \sum_{i=1}^{4}( E  N_{i} +   U\, N_{i \uparrow}  N_{i \downarrow})
 \end{eqnarray}
Operators and variables (except $t$) in lower/upper case relate to the
ligands/metal ion and  all indices
run from 1 to 4 cyclically. $\langle i,j\rangle$ denotes a summation over
neighboring metal ions $i$ and ligands $j$. The operator $a^{\dag}_{j \sigma}$ creates
 an electron on ligand site $j$ with spin $\sigma$, $n_{j \sigma} =  a^{\dag}_{j \sigma} a_{j \sigma}$ and $n_{j} =
\sum_{\sigma} n_{j \sigma}$. Similar definitions hold
for the metal ions: $ A_{i \sigma}, N_{i \sigma} = A^{\dag}_{i
  \sigma} A_{i \sigma},N_{i}=\sum_{\sigma} N_{i \sigma}$.
 The tunneling term (\ref{eq:ham_tun}) describes hopping between
 ligand and metal ions and is assumed to be independent of $i$ and $j$ due to
 molecular symmetry. The ligand-part of the Hamiltonian (\ref{eq:ham_lig}) consists of an orbital with energy $\epsilon$, the
Coulomb repulsion terms on the ligand ($u$) and between adjacent ($v$) and opposite ligands ($w$).  Due to decreasing overlap
with distance we have $u > v > w$.  Hamiltonian (\ref{eq:ham_met})
describes the isolated metal ion orbitals with energy $E$. Here we only consider the short-range interaction $U$
because the $d$  orbital overlap between two ions is typically much smaller than
that between two ligand orbitals. In Fig.~\ref{fig:geometry} these
interactions are schematically indicated. We study the parameter regime where the first eight extra \textit{electrons
occupy} four equivalent \textit{ligand centered orbitals}. Such a sequence has been well-documented for a number of grid-molecules~\cite{ruben03a}.
In our model this is the case when the charge excitations of the ligand lie below
the ones of the  metal ion: $ \epsilon < \epsilon +u < E < E+U$.
The metal ligand charge-transfer barrier $\Delta = \epsilon -E  $ suppresses the direct hopping of extra electrons from the ligands to
 unoccupied metal ions:  $|\Delta| \gg t$. The
fluctuations of the metal orbital occupation around zero can be treated
using  a Schrieffer-Wolff transformation~\cite{schrieffer66}.  We obtain
an extended Hubbard model $H_{\mathsf{eff}} =
\sum_{\langle j k \rangle}\sum_{\sigma} t_{\mathsf{eff}}
\,  a^{\dag}_{j \sigma} a_{k \sigma} + H_{\mathsf{L}}$ on four
ligand sites with an effective  hopping matrix element $t_{\mathsf{eff}}=-\frac{t^2}{2 \Delta}$.\\
\emph{Addition energies and spin states ---}
We first analyze the electronic spectrum of
 $H_{\mathsf{eff}}$ for different
charge sectors. In the case $u > v > w/2 >
t_{\mathsf{eff}}$ the first electron reduces one of the four
ligands, say $j=1$. The next one occupies the opposite ligand $j=3$ in
 order to minimize the Coulomb interaction. The next electrons reduce the adjacent ligands $j=2,4$.
 This sequence is repeated for the next four electrons, each time doubly occupying a ligand orbital. 
 The gaps in the addition spectrum $w, 2v-w,w, u, w,2v-w,w$ ({\em
 extra} energy required for the next electron) thus directly relate to
 geometrical features of the grid~\cite{ruben03a}. These
  electrostatic parameters correspond directly to the size of the
  Coulomb diamonds obtained in transport experiments.
\begin{figure}
\includegraphics[scale=0.6]{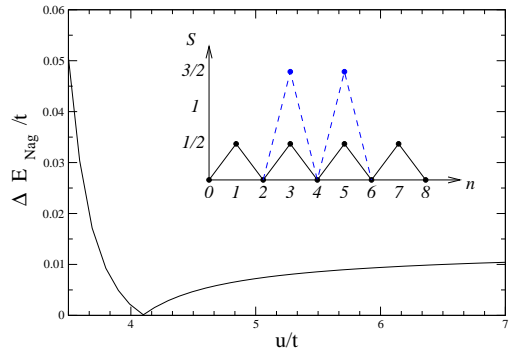}
\caption{\label{fig:split_combined} Splitting $\Delta E_{\mathsf{Nag}}$ for $n=3$ as function of $u$ for $
  v=2.25 t, w=1 t, \Delta =-10 t, t_{\mathsf{eff}}=0.05 t$. For $u_{\mathsf{th}} \approx
  4.15 t= 83 t_{\mathsf{eff}}$ the ground state has maximal
  spin. Inset: ground state spin as function of the
  number of electrons $n$ added to the ligands for
  $u<u_{\mathsf{th}}$ (solid black line) and  $u>u_{\mathsf{th}}$ (dashed blue line).}
\end{figure}
Now we discuss the ground state spin as successive electrons are added
to the molecule. Due to superexchange processes electron spins on neighboring ligands
tend to couple antiferromagnetically. This leads to an alternating sequence of
$S=0$ and $S=1/2$ as electrons are added (inset Fig.~\ref{fig:split_combined}). However, for \textit{sufficiently large}  $u > u_{\mathsf{th}}$ and
fixed $t$, resp. $t_{\mathsf{eff}}$ (Fig.~\ref{fig:split_combined}) the ground state spin for $n=3,5$ is enhanced from the noninteracting
value $S=1/2$ to the maximal
possible value $S=3/2$.
Because double occupation is suppressed, a single hole/electron (relative to the
half-filled state $n=4$) can maximally gain kinetic energy when the
background of the other electrons is fully spin polarized.
 This is the underlying mechanism for the Nagaoka
theorem~\cite{nagaoka66}.
 The ferromagnetic alignment (due to complete delocalization) competes with
the antiferromagnetic spin coupling (due to superexchange processes
between neighboring occupied sites). Which process dominates
depends on the strength of the onsite repulsion $u$ relative to fixed
hopping.
The  interactions $v,w$ tend to increase the threshold value $u_{\mathsf{th}}$ for fixed $t_{\mathsf{eff}}$~\cite{kollar96}.
The gap between the Nagaoka state and the lowest excited state
saturates at  $\Delta E_{\mathsf{Nag}} \sim 2 t_{\mathsf{eff}}$  (Fig.~\ref{fig:split_combined}) \emph{independent} of $u$ due to the
kinetic origin of the effect.
 In order to
attain an observable effect one should thus have
sufficiently large values of both $u$ \textit{and} $t_{\mathsf{eff}}$.
For example, $u$ can be increased by a chemical modification of the ligands which draws charge density into the ligand
LUMO orbitals. Taking typical
parameters~\cite{ruben03a} $ |\Delta| \approx 1 eV, t \approx
10^{-1} eV, \Delta E_{\mathsf{Nag}} \approx 10^{-2}eV, $ we estimate $ u_{\mathsf{th}} \approx
1 eV $ which is reasonable.
\\

\textbf{Transport---}
The charge sensitivity of the total spin
 is observable in the
single electron  tunneling-current through the molecule.
 To demonstrate this, we consider the  Hamiltonian
$ H = H_{\mathsf{res}} + H_{\mathsf{eff}} + H_{\mathsf{mol-res}}$, employing
 units $\hbar = e =k_{B} = 1$. The electrodes $r=L,R$  are described as electron
reservoirs with electrochemical potentials $\mu_{r}=\mu
\pm V/2$ and a constant density of states $\rho$:
$H_{\mathsf{res}}= \sum_{k \sigma r} \epsilon_{k \sigma r} c^{\dag}_{k \sigma r} c_{k \sigma r}$.
The tunneling term $H_{\mathsf{mol-res}} =  (\frac{\Gamma}{2 \pi \rho})^{1/2} \sum_{k \sigma j r}
t_j^r c^{\dag}_{k \sigma r} a_{j \sigma} + h.c. $ describes
charge transfer between electrode and molecule (symmetric tunneling barriers). $\Gamma$ is the overall
coupling strength between leads and the molecule and defines the current scale. We assume that
tunneling is only possible through two ``contact'' ligands, namely
$t_1^L=t_3^R=1$, otherwise $0$. We have checked that this choice does not cause effects due
to orbital symmetry as discussed in~\cite{hettler03} by trying also $j=1,2$.
The coupling to a gate electrode is included in a shift of the
single particle energies $\epsilon \rightarrow \epsilon
-\alpha V_{\mathsf{g}}$.
\begin{figure}
\includegraphics[scale =0.4]{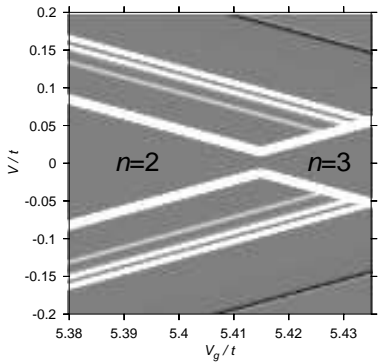}
\caption{\label{fig:sblock}  $dI/dV(V_{\mathsf{g}}, V)$ grayscale plot
    (white/black~$\gtrless 0$)
    for $u=5t, v=2.25t, w=t, \Delta=-10t, T=2~10^{-4}t$.}
\end{figure}
In the weak tunneling regime ($\Gamma \ll T$) the effect of the leads can be
incorporated in the transition rates $ \Sigma_{s,s'} = \sum_{r}  \Sigma_{s,s'}^{r,+}+\Sigma_{s,s'}^{r,-}$ between the molecular many-body
states $s,s'$:
 \begin{eqnarray}
  \label{eq:rates}
  \Sigma_{s,s'}^{r,+}& = & \Gamma  \sum_{ \sigma}    f_{r}^{+}(E_{s}-E_{s'})
  | \sum_{j} t_j^r  \langle  s |  a^{\dag}_{j \sigma} |  s' \rangle |^{2} \nonumber \\
 \Sigma_{s,s'}^{r,-}& = & \Gamma \sum_{\sigma} f_{r}^{-}(E_{s}-E_{s'}) | \sum_{j} t_j^r \langle  s | a_{j \sigma} |  s' \rangle |^{2}.
\end{eqnarray}
Here $f_{r}^{+}$ is the Fermi function of reservoir $r$ and  $f_{r}^{-}=1-f_{r}^{+}$.
Importantly, the matrix elements include the calculated many-body wavefunction of the molecule and the spin selection rules. From the stationary master equation $
 \sum_{s'} (\Sigma_{s,s'}P_{s'}-\Sigma_{s',s}P_{s})=0 $
we obtain the non-equilibrium occupations $P_s$ of the molecular states $s$ and the resulting stationary current
which may be calculated at either electrode $r=L,R$:
\begin{eqnarray}
  \label{eq:cur}
  I_r  = -  \sum_{s, s'}(\Sigma_{s,s'}^{r,+}
  P_{s'}-\Sigma_{s',s}^{r,-}P_{s}).
\end{eqnarray}
Due to the presence of a maximal spin state, either as ground or
excited state, spin blockade and  NDC effects occur, respectively~\cite{weinmann95}.\\
\emph{Maximal spin ground state ---}
For $u > u_{\mathsf{th}}$ and fixed $t$ the \emph{ground state} for
$n=3,5$ is a Nagaoka state (Fig.~\ref{fig:split_combined}).
Transport involving groundstates  $n \leftrightarrow n+1$ for $n=2,
\cdots, 5$ is completely blocked for
\textit{small bias} and low temperature ($T,V<\Delta E_{\mathsf{Nag}}$): the
Coulomb diamonds in a differential conductance versus $(V_g,V)$ plot
do not close (Fig.~\ref{fig:sblock}).
Since the ground state spin is either  $0$ ($n=2,4,6$) or $3/2$ ($n=3,5$) the tunneling rates between neighboring ground states vanish:
even when ground state transitions are energetically allowed, due
to the spin selection rule  $\Delta S = 1/2 $  transport is completely
blocked in the weak tunneling limit. However, when temperature or voltage are increased, such that
 the first excited state with appropriate spin can be accessed, current begins to flow.
\begin{figure}[t]
\includegraphics[scale =0.4]{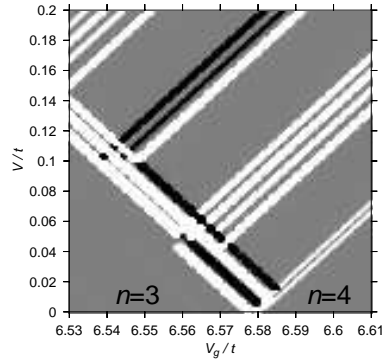}
\caption{\label{fig:ndc}  Same parameters as Fig.~\ref{fig:sblock} except $u=4.05t$ and
    around a different charge degeneracy point.}
\end{figure}
\newline
\emph{Maximal-spin excited state ---}
Depending on the gate voltage, NDC and even complete current
 blocking can occur at \textit{finite bias voltage}~\cite{weinmann95}  when the
 Nagaoka state is the lowest spin-excitation for $n=3,5$ (i.e. $u
 \lesssim u_{\mathsf{th}}$, Fig.\ref{fig:split_combined}).
 The typical result is depicted in Fig.~\ref{fig:ndc}.
 Near the charge $3 \leftrightarrow 4$ degeneracy point (and $4
 \leftrightarrow 5$) two NDC lines (black) with
 negative slope appear which are due to the low lying $S=3/2$ Nagaoka
 state.
 We first discuss  the lower NDC line using the left scheme in
 Fig.~\ref{fig:level}. At the charge degeneracy point ($V_g/t \sim
 6.58$ in Fig.~\ref{fig:ndc}) the current sets on because a transport channel is opened,
 namely $n=3, S=1/2$~(2-fold degenerate)~$ \leftrightarrow n=4, S=0$. Increasing the bias voltage
 by $\Delta_4$ results in a gain in population of the Nagaoka state via the $n=4, S=1$ excited
 state. Since the Nagaoka state cannot decay
 (strictly) to the $n=4, S=0$ ground state the number of transport
 channels is therefore decreased from two to one leading to the lower NDC
 effect. Further away from the degeneracy point ($V_g/t \lesssim 6.56$ in
Fig.~\ref{fig:ndc}), the lower NDC line turns into a conductance peak
and simultaneously the ground-state transition line below it
disappears. This is due to a complete population inversion
between the ground and excited state for $n=3$ which already occurs inside the Coulomb diamond. 
When the transition from $n=4, S=0$ to the third excited state $n=3,
S=1/2$ lies in the bias window,
the Nagaoka state is occupied starting from the ground-state
$n=3,S=1/2$ via the cascade of single-electron tunneling processes
indicated in the right panel of Fig.~\ref{fig:level}.
It is \textit{fully} occupied because the escape rate from $n=3,
S=3/2$ relative to that from $n=3, S=1/2$ ground-state
is suppressed by a factor $\sim e^{-(\Delta_4-\Delta_3) / T} \ll 1$ (Fig.~\ref{fig:level}).
 \begin{figure}
\includegraphics[scale =0.33, angle=-90]{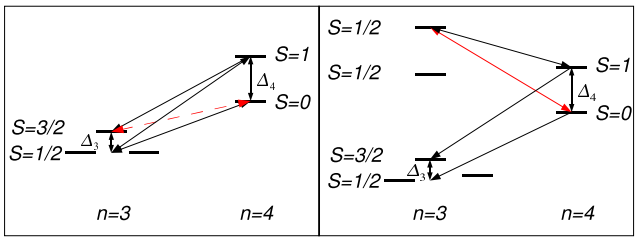}
\caption{\label{fig:level} Minimal set of states for the lower NDC effect (left) and the complete current 
suppression (right) in Fig.~\ref{fig:ndc}.}
\end{figure}
 The upper NDC line with negative slope in Fig.~\ref{fig:ndc} is caused by the occupation of the high lying
 maximal-spin state $n=4,S=2$ which cannot decay to states with one
 electron less and {higher} spin~\cite{weinmann95}. This state can
 already be reached at low voltages only due the presence of the low-lying Nagaoka state $S=3/2$ at $n=3$.
\\

\textbf{Conclusion---}
 We have shown that in particular types of single-molecule devices
 the Nagaoka spin-polarization mechanism
 is relevant since the Coulomb blockade
effect allows the controlled addition of single electrons (in contrast to the
case of bulk magnets). The magnetic properties of the molecule may thus be switched by
a gate voltage. The fingerprints of this charge-selective stabilization of  maximal
spin states (either a ground or a low-lying excited state)
 are observable in the  tunneling current.
We also have investigated an extension of the model considered here
with a spin degree of freedom added to each of the four metal ions~\cite{romeike04}.
In this more complicated case  the cooperative spin-polarization effect (of both Nagaoka-origin and
 direct ion-ligand exchange) compete with
the antiferromagnetic superexchange (induced by ion-ligand  hopping).
It is found that the Nagaoka state is also relevant here
and that its transport fingerprints are
the same as demonstrated in this work for a generic model.
\\

We thank M. Ruben and J. Kortus for stimulating discussions. M. R. W. acknowledges the financial support provided through the European
Community's Research Training Networks Program under contract HPRN-CT-2002-00302, Spintronics.

\end{document}